\documentclass[%
preprint,
 amsmath,amssymb,
 aps
]{revtex4-1}

\usepackage{graphicx}
\usepackage{dcolumn}
\usepackage{bm}
\usepackage[colorlinks=true,citecolor=blue,linkcolor=blue,urlcolor=blue]{hyperref}

\def\be{\begin{equation}}
\def\ee{\end{equation}}

\begin{document}


\title{The speed of gravitational waves and power-law solutions in a scalar-tensor model }

\author{L. N. Granda}
 \email{luis.granda@correounivalle.edu.co} 
\author{D. F. Jimenez}%
 \email{jimenez.diego@correounivalle.edu.co}
\affiliation{%
Departamento de Fisica, Universidad del Valle
\\
A.A. 25360, Cali, Colombia 
}%


%
%

\begin{abstract}
One of the most relevant solutions in any cosmological model concerning the evolution of the universe is the power-law solution. For the scalar-tensor model of dark energy with kinetic and Gauss Bonnet couplings, it is shown that we can conserve the power-law solution and at the same time meet the recent observational bound on the speed of gravitational waves. In the FRW background the anomalous contribution to the speed of gravitational waves, coming from the kinetic and Gauss-Bonnet couplings, cancel each other for power-law solutions. It is shown that by simple restriction on the model parameters we can achieve a non-time-dependent cancellation of the defect in the velocity of the gravitational waves.
The model can realize the cosmic expansion with contributions from the kinetic and Gauss-Bonnet couplings of the order of ${\cal O}(1)$ to the dark energy density parameter. The results are valid on the homogeneous FRW background and the limitations of the approach are discussed.
\begin{description}
\item[PACS numbers]
04.50.-w, 04.50.kd, 95.36+x
\end{description}
\end{abstract}

\maketitle

\section{\label{intro}introduction}
The recent detection of gravitational waves (GW) from the merger of neutron stars in the system GW170817 together with its electromagnetic counterpart, the gamma-ray burst GRB 170817A \cite{gw1, gw2,gw3}, imposes strong bound on the speed of GW $c_g$ which should differ from the speed of light $c$ by at least one part in $10^{15}$. At the same time this bound on the speed of GW translates into strong constraints on a widely studied scalar-tensor theories of gravity \cite {horndeski1,dgsz} which have successfully explained some of the aspects of the current dark energy problem (see reviews \cite{copeland, amendola, clifton, nojiri,nojiri1}). 
Horndeski theories are the most general scalar-field theories with equations of motion with no higher than second order derivatives. They include the known dark energy models of quintessence, k-essence, non-minimal coupling of the scalar field to curvature, to the Gauss-Bonnet invariant and non-minimal kinetic couplings to curvature, among other interactions. 
The interaction terms containing the coupling of the scalar field to the Gauss-Bonnet invariant, and the coupling between the scalar kinetic term and the curvature appear, among others,  in the $\alpha'$-expansion of the string effective action \cite{metsaev, meissner}. The non-minimal couplings of the kinetic term to curvature were considered in \cite{amendola1, scapozziello} to study inflationary attractor solutions, in \cite{scapozziello1} to find a connection with the cosmological constant, and a variety of solutions for the different cosmological epochs, particularly for late time acceleration, were found in \cite{vsushkov}-\cite{sushkov4}, \cite{gubitosi},  \cite{granda1},  {\cite{granda2}. The coupling between the scalar field and the Gauss-Bonnet invariant has been  proposed to address the dark energy problem in \cite{sergei4}, where it was found that quintessence or phantom phase may occur in the late time universe. Different aspects of accelerating cosmologies with GB correction have been also discussed in \cite{tsujikawa1}, \cite{leith}, \cite{koivisto1}, \cite{koivisto2}, \cite{neupane2}, and a modified GB theory applied to dark energy have been suggested in \cite{nojiriod1}. Late time cosmological solutions in a model that includes both, non-minimal kinetic coupling to curvature and coupling of the scalar field to the Gauss-Bonnet invariant, have been studied in \cite{granda5}-\cite{granda7}.\\
A very important feature of these general scalar-tensor models is that they predict an anomalous GW speed ($c_g\ne c$), entering in contradiction with the observed results from GW170817 and GRB 170817A, and therefore are serious candidates to be discarded as dark energy models.
The implications that this discovery has for the nature of dark energy (DE), apart form the tests of General Relativity, have been highlighted in \cite{lucas1,lucas2, bettoni, creminelli, jose, bellini}. Constraints imposed by the speed of GW on a scalar-tensor theories were analyzed in \cite{jose, nojodin, jeremy, simone, langlois}. In \cite{lamendola1} it was shown that the scalar field should be conformally coupled to the curvature in order to avoid the restrictions imposed by the speed of GW. Restrictions on vector-tensor theories have also been analyzed in \cite{bellini, lamendola1}. The GW constraints on the coupling of the scalar kinetic term to the Einstein tensor and the scalar field to the GB invariant have been studied in \cite{gong}, and restrictions on the beyond Horndeski parameters have been analyzed in \cite{crisostomi,dima}.  According to all above studies, the overall conclusion is that the constraint on $\alpha_g$ leaves only models that are conformally coupled to gravity, including models where the gravity is minimally coupled and models such as $f(R)$ gravity. A model independent phenomenological analysis of Horndeski theory on FRW background is given in \cite{arai}, where it was shown that Horndeski theory with arbitrary $G_4$ and $G_5$ can hardly account for $c_g=c$ and explain the accelerating expansion without fine-tuning. 
Despite the severe restrictions imposed by GRB 170817A, in this work we show that there is an important solution, namely the power-law expansion in the frame of a scalar-tensor model with non-minimal kinetic and Gauss-Bonnet couplings, that passes the test of the speed of GW and remains consistent, at least in the FRW background, even after GRB 170817A. The relevance of the power-law solutions lies in the fact that they describe the different asymptotic regimes of expansion, depending on the type of matter that dominates throughout the evolution of the universe. We show that the cancelation between the anomalous contributions to $c_g$ coming from the kinetic and GB couplings is not time-dependent and the model parameters don't require to be tuned beyond what is necessary to comply with the dark energy observations.\\
The limitation of the present approach lies in fact that we used $\dot{phi}$, $\ddot{\phi}$ and $H$, defined on the homogeneous FRW background, in the cancellation of the GW speed anomaly. In other words, the cancelation between the contributions from the interaction Lagrangians occurs at the cost of "tuning" the background manifold, while a definitive cancelation of the GW speed anomaly demands a covariant approach which makes it independent of the background metric. A discussion on the covariant approach is given.
The paper is organized as follows. In section II we give the general equations expanded on the FRW metric, and find the power-law solutions for quintessence-like and phantom expansion. In section III we give the speed of GW for the model on the homogeneous FRW background, and find the restrictions on the parameters that cancel the anomalous contribution to  the speed of GW. In section IV we give a summary and discussion.
\section{The model and power-law solutions}
We consider the following string motivated action which includes the Gauss Bonnet coupling to the scalar field and kinetic couplings to curvature. These terms are present in the next to leading $\alpha'$ corrections in the string effective action (where the coupling coefficients are functions of the scalar field) \cite{metsaev}, \cite{meissner} . 
\be\label{eq1}
\begin{aligned}
S=&\int d^{4}x\sqrt{-g}\Big[\frac{1}{16\pi G} R -\frac{1}{2}\partial_{\mu}\phi\partial^{\mu}\phi+F_1(\phi)G_{\mu\nu}\partial^{\mu}\phi\partial^{\nu}\phi\\&- V(\phi)-F_2(\phi){\cal G}\Big]
\end{aligned}
\ee
\noindent where $G_{\mu\nu}=R_{\mu\nu}-\frac{1}{2}g_{\mu\nu}R$, ${\cal G}$ is the 4-dimensional GB invariant ${\cal G}=R^2-4R_{\mu\nu}R^{\mu\nu}+R_{\mu\nu\rho\sigma}R^{\mu\nu\rho\sigma}$. The coupling $F_1(\phi)$ has dimension of $(length)^2$, and the coupling $F_2(\phi)$ is dimensionless. 
Note that we are not considering derivative terms that are not directly coupled to curvature, of the form $\Box\phi\partial_{\mu}\phi\partial^{\mu}\phi$ and $(\partial_{\mu}\phi\partial^{\mu}\phi)^2$. The equations derived from this action contain only second derivatives of the metric and the scalar field.\\
In the spatially-flat Friedmann-Robertson-Walker (FRW) metric,
\be\label{eq2}
ds^2=-dt^2+a(t)^2\left(dr^2+r^2d\Omega^2\right),
\ee
where $a(t)$ is the scale factor, the set of equations describing the dynamical evolution of the FRW background and the scalar field for the model (\ref{eq1}) are ($8\pi G=\kappa^2=M_p^{-2}$)
\be\label{eq3}
H^2=\frac{\kappa^2}{3}\left(\frac{1}{2}\dot{\phi}^2+V(\phi)+9 H^2F_1(\phi)\dot{\phi}^2+24H^3\frac{dF_2}{d\phi}\dot{\phi}\right)
\ee
\be\label{eq4}
\begin{aligned}
&\ddot{\phi}+3H\dot{\phi}+\frac{dV}{d\phi}+3 H^2\left(2F_1(\phi)\ddot{\phi}+\frac{dF_1}{d\phi}\dot{\phi}^2\right)+
\\&18 H^3F_1(\phi)\dot{\phi}+12 H\dot{H}F_1(\phi)\dot{\phi}+24\left(\dot{H}H^2+H^4\right)\frac{dF_2}{d\phi}=0
\end{aligned}
\ee
Here we will consider the string inspired model with exponential couplings, and additionally, we consider an exponential potential given by
\be\label{eq5}
\begin{aligned}
F_1(\phi)=&\xi e^{\alpha\kappa\phi/\sqrt{2}},\,\,\,\,  F_2(\phi)=\eta e^{\alpha\kappa\phi/\sqrt{2}},\\& V(\phi)=V_0 e^{-\alpha\kappa\phi/\sqrt{2}}
\end{aligned}
\ee
Where the coupling $\eta$ may be related to the string coupling $g_s$ as $\eta\sim 1/g_s^2$. The de Sitter solution for the model (\ref{eq1}) with couplings and potential given by (\ref{eq5}) follows from Eqs. (\ref{eq3}) and (\ref{eq4}) by setting $H=const.=H_c$ and $\phi=const.=\phi_c$, which gives
\be\label{eq5a}
H_0^2=\frac{M_p^2}{8\eta}e^{-2\alpha\kappa\phi_c/\sqrt{2}} 
\ee
This model admits the power-law solution for quintessence-like expansion \cite{grajisa,gralo}
\be\label{eq7}
H=\frac{p}{t},\,\,\,\,\, \phi=\phi_0\ln\frac{t}{t_1},
\ee
and the solution
\be\label{eq8}
H=\frac{p}{t_s-t},\,\,\,\,\, \phi=\phi_0\ln\frac{t_s-t}{t_1}
\ee
for phantom power-law expansion. These solutions lead to the effective equation of state $w_{eff}$ 
\be\label{eq8a}
w_{eff}=-1\pm \frac{2}{3p}
\ee
where the lower sign is for phantom solution.
The equations (\ref{eq7}) and (\ref{eq8}), after being replaced in the Friedmann equation (\ref{eq3}), lead to the following restrictions
\be\label{eq9}
\frac{3p^2}{\kappa^2}=\frac{1}{2}\phi_0^2+V_0 t_1^2+\frac{9\xi p^2}{t_1^2}\phi_0^2\pm \frac{48\eta p^3}{t_1^2}
\ee
where we fixed $\phi_0$ according to
\be\label{eq10}
\frac{2\sqrt{2}}{\alpha\kappa}=\phi_0,
\ee
in order to get  the power $t^{-2}$ from the interacting terms. The equation of motion (\ref{eq4}) gives
\be\label{eq11}
(\pm 3p-1)\phi_0^2-2V_0 t_1^2+\frac{6\xi p^2(\pm 3p-2)}{t_1^2}\phi_0^2\pm\frac{48\eta p^3(\pm p-1)}{t_1^2}=0
\ee
where the lower minus sign follows for the phantom solution. The stability properties of the solutions (\ref{eq7}) and (\ref{eq8}) has been performed in  \cite{grajisa} for the case of $V_0=0$ and in \cite{gralo} for more general cases. As will be shown below, the restrictions (\ref{eq9})-(\ref{eq11}) with an additional constraint coming from the velocity of the GW can be solved consistently.\\
\section{Restriction from the speed of gravitational waves}
The generalized Galileons, which are equivalent to  Horndeski theory in four dimensions, represent the most general scalar field theories having second-order field equations and are described by the Lagrangian density \cite{dgsz, kobayashi}
\be\label{eq12}
{\cal L}=\sum_{i=2}^5 {\cal L}_i
\ee
with
\be\label{eq13}
{\cal L}_2=K(\phi,X),\;\;\; {\cal L}_3=-G_3(\phi,X)\Box \phi,
\ee
\be\label{eq15}
{\cal L}_4=G_4(\phi,X)R+G_{4,X}\left[(\Box \phi)^2-(\nabla_{\mu}\nabla_{\nu}\phi)^2\right],
\ee
\be\label{eq16}
\begin{aligned}
{\cal L}_5=&G_5(\phi,X)G_{\mu\nu}\nabla^{\mu}\nabla^{\nu}\phi-\frac{G_{5,X}}{6}\Big[(\Box\phi)^3-\\&3\Box\phi(\nabla_{\mu}\nabla_{\nu}\phi)^2+2(\nabla_{\mu}\nabla_{\nu}\phi)^3\Big]
\end{aligned}
\ee
where $X=-\frac{1}{2}\nabla_{\mu}\phi\nabla^{\mu}\phi$, $G_{\mu\nu}$ is the Einstein tensor and $G_{i,X}=\partial G_i/\partial X$.\\
The Einstein general relativity takes place for $K=G_3=G_5=0$ and $G_4=M_p^2/2$. To identify to which the sector of the generalized Galileons, the Lagrangian density (\ref{eq1}) is related, we use the following correspondence  \cite{kobayashi, copeland1}:\\
the terms
\be\label{eq17}
\frac{R}{2\kappa^2}+F_1(\phi)G_{\mu\nu}\partial^{\mu}\phi\partial^{\nu}\phi
\ee
generate the Galilean functions \cite{copeland1}
\be\label{eq18}
K=2F_1''(\phi)X^2,\;\; G_3=3F_1'(\phi)X,\;\;\; G_4=\frac{M_p^2}{2}+F_1(\phi)X
\ee
and the Gauss-Bonnet coupling $F_2(\phi){\cal G}$ generates \cite{kobayashi, copeland1}
\be\nonumber
K=-8F_2''''(\phi)X^2\left(3-\ln(X)\right),\; G_3=-4F_2'''(\phi)X\left(7-3\ln(X)\right)
\ee
\be\label{eq19}
G_4=-4F_2''(\phi)X\left(2-\ln(X)\right),\;\;\; G_5=4F_2'(\phi)\ln(X)
\ee
where $'$ denotes derivative with respect to $\phi$ and $X=-1/2\partial_{\mu}\phi\partial^{\nu}\phi$. Thus, the explicit functions $K$ and $G_i$ that establish the correspondence of the model (\ref{eq1}) with the general Galileons are given by 
\be\label{eq20}
K=2F_1''(\phi)X^2-8F_2''''(\phi)X^2\left(3-\ln(X)\right)
\ee
\be\label{eq21}
G_3=3F_1'(\phi)X-4F_2'''(\phi)X\left(7-3\ln(X)\right)
\ee
\be\label{eq22}
G_4=\frac{M_p^2}{2}+F_1(\phi)X-4F_2''(\phi)X\left(2-\ln(X)\right)
\ee
\be\label{eq23}
G_5=4F_2'(\phi)\ln(X)
\ee
Then, the generalized Galileons Lagrangian density equivalent to the one of the model (\ref{eq1}) has the explicit form
\begin{widetext}
\be\label{eq23o}
\begin{aligned}
{\cal L}=&2F_1''(\phi)X^2-8F_2''''(\phi)X^2\left(3-\ln(X)\right)-\Big[3F_1'(\phi)X-4F_2'''(\phi)X\left(7-3\ln(X)\right)\Big]\Box\phi+\\&
\left[\frac{M_p^2}{2}+F_1(\phi)X-4F_2''(\phi)X\left(2-\ln(X)\right)\right]R+\Big[F_1-4F_2''\left(1-\ln(X)\right)\Big]\Big[(\Box\phi)^2-(\nabla_{\mu}\nabla_{\nu}\phi)^2\Big]\\& +4F_2'(\ln(X))G_{\mu\nu}\nabla^{\mu}\nabla^{\nu}\phi-\frac{2}{3}X^{-1}F_2'\Big[(\Box\phi)^3-3\Box\phi(\nabla_{\mu}\nabla_{\nu}\phi)^2+2(\nabla_{\mu}\nabla_{\nu}\phi)^3\Big]
\end{aligned}
\ee
\end{widetext}
The second order action for the tensor perturbations $h_{ij}$, which generate the GW, can be written in the form  \cite{kobayashi, felice, bellini1}. 
\be\label{eq23a}
S^{(2)}_h=\frac{1}{8}\int dtd^3x a^3\left[{\cal Q}_h \dot{h}_{ij}^2-\frac{{\cal F}_h}{a^2}(\nabla h_{ij})^2\right],
\ee
where
\be\label{eq23b}
{\cal Q}_h=2\left[G_4-2XG_{4X}- X\left(\dot{\phi}HG_{5X}-G_{5\phi}\right)\right],
\ee
\be\label{eq23c}
{\cal F}_h=2\left[G_4-X\left(\ddot{\phi}G_{5X}+G_{5\phi}\right)\right]
\ee
which gives the squared speed of propagation for the GW as
\be\label{eq23d}
c_g^2=\frac{{\cal F}_h}{{\cal Q}_h}.
\ee
It is convenient to write $c_g^2$ in terms of the velocity of light ($c=1$) plus an anomalous contribution, $\alpha_g$, which reflects the contribution to $G_4$ and $G_5$ coming from the non minimally coupled terms
\be\label{eq23e}
c_g^2=1+\alpha_g
\ee
In order to satisfy the GW170817 and GRB 170817A observations, it is natural to consider $\alpha_g\simeq 0$ and try to analyze its implications for the corresponding model. From (\ref{eq23b})-(\ref{eq23d}) one find the following expression
\be\label{eq24}
\alpha_g=\frac{X\left[2G_{4,X}-2G_{5,\phi}-(\ddot{\phi}-\dot{\phi}H)G_{5,X}\right]}{G_4-2XG_{4,X}+XG_{5,\phi}-\dot{\phi}HXG_{5,X}}
\ee
Note that for the minimally coupled models, the cancelation of $\alpha_g$ is trivial, while for the non-zero terms $G_{4X}$, $G_{5\phi}$ and $G_{5X}$ it has not been possible to achieve $\alpha_g\sim 0$ without finely-tuned cancellations \cite{jose, bellini, nojodin, jeremy, simone, langlois, lamendola1}. However, as shown bellow, there are important cosmological solutions  corresponding to asymptotic regimes of cosmological expansion, that pass the test of the GW170817 and GRB 170817A observations, even in the frame of scalar-tensor models.
For the present model the contribution to $\alpha_g$ comes from the interactions related to the kinetic and GB couplings.
These anomalies can be obtained from the model (\ref{eq1}), using the correspondence equations for $G_4$ (\ref{eq22}) and $G_5$ (\ref{eq23}), which give
\be\label{eq25}
\alpha_g=\frac{2X\left(F_1-4F_2''\right)-4\left(\ddot{\phi}-\dot{\phi}H\right)F_2'}{M_p^2/2-XF_1-4\dot{\phi}HF_2'}
\ee
In order to satisfy the observed restriction on the velocity of GW (i.e. $\alpha_g<10^{-15}$), we require $\alpha_g=0$ and look for the conditions to cancel de numerator in (\ref{eq25}). First we note that the de Sitter solution (where $\phi=const.$) leads automatically to $\alpha_g=0$. By replacing the couplings $F_1$ and $F_2$ from (\ref{eq5}) and noticing that the exponential is a common factor to the whole expression, we get 
\be\label{eq26}
\left(\xi-4\eta\left(\frac{\alpha\kappa}{\sqrt{2}}\right)^2\right)\dot{\phi}^2-4\eta\left(\ddot{\phi}-\dot{\phi}H\right)\frac{\alpha\kappa}{\sqrt{2}}=0,
\ee
This is the condition for the cancelation of the  anomalous contribution to $c_g$, from the non-minimal kinetic and GB couplings, when these couplings have the exponential form (\ref{eq5}). Note that this constraint has a general character in the sense that it can be used to restrict the dynamics of the model, and it is not just a fine tune.   
Applying the restriction (\ref{eq26}) to the power-law solutions (\ref{eq8}) and (\ref{eq9}) we can see that the time dependence can be factorized from the whole expression, and (\ref{eq26}) leads to
\be\label{eq27}
\xi\phi_0^2+8\eta(\pm p-1)=0
 \ee
where we used (\ref{eq10}). This restriction establishes a relationship between the parameters $\eta$, $\xi$, $\alpha$ and the nature of the expansion encoded in $p$. It should be noted that although we have involved $\dot{\phi}$, $\ddot{\phi}$ and $H$, which are related to the FRW background, in the cancellation of the GW speed anomaly, the resulting expression (\ref{eq27}) does not depend on the FRW coordinates. Hence, if this restriction takes place, then $\alpha_g=0$ during all the time evolution in an homogeneous FRW background.
With this restriction added to the equations (\ref{eq9})-(\ref{eq11}), and solving the resulting system with respect to $V_0$, $\xi$ and $\eta$ one finds
\be\label{eq31}
V_0=\frac{12p^2(p\mp 1)^2M_p^2+(p^2\mp 6p+1)\phi_0^2}{2(2p^2\mp 3p-1)t_1^2}
\ee
\be\label{eq32}
\xi=\frac{(p\mp 1)(\pm 2pM_p^2-\phi_0^2)t_1^2}{2p(2p^2\mp 3p-1)\phi_0^2}
\ee
\be\label{eq33}
\eta=\frac{(\pm \phi_0^2-2pM_p^2)t_1^2}{16p(2p^2\mp 3p-1)}
\ee
where the upper sign is for quintessence-like and the lower is for phantom solutions. Note that the potential is always positive for the values of $p$ of interest for accelerated expansion (for instance $p\sim 20$ gives $w_{eff}\sim -1\pm 0.03$). In the limit $\eta\rightarrow 0$, the restriction (\ref{eq27}) leads to $\xi\rightarrow 0$ giving the standard uncoupled scalar field without the anomaly in the propagation of GW. 
Note that from the expression for $H(t)$ we can find a useful appropriate value for $t_1$ if we assume the initial condition $H(t_1)=H_0$, giving   $t_1=pH_0^{-1}$ or $H=H_0t_1/t$. Then, it also follows from (\ref{eq31})-(\ref{eq33}) that at $\phi_0=\sqrt{2p}M_p$, it is obtained 
$$\xi=0,\;\;\; \eta=0,\;\;\;\; V_0=\frac{(3p-1)}{p}M_p^2H_0^2,$$ 
which reproduces exactly the power-law solution for the quintessence uncoupled scalar field with exponential potential.
In our case, by taking $\phi_0=pM_p/4$ and using the initial condition $H(t_1)=H_0$, we can give an estimate of the contribution of the potential and of each coupling to the dark energy density, in order to see if they are relevant for the cosmic accelerated expansion (in addition to satisfying the constraint in $c_g$). Taking $p=20$, one finds for quintessence-like expansion (upper sign), that the potential gives a contribution $\sim 0.98 H_0^2M_p^2$, the kinetic term contributes $\sim 0.46 H_0^2M_p^2$ and the GB term contributes $\sim 0.41 M_p^2H_0^2$, being all consistent with what should be expected, that is $\sim {\cal O}(1) M_p^2H_0^2$. 
It is worth noting that the effective Planck mass $M_p^*$ defined as
\be\label{eq33b}
M_p^{*2}=M_p^2-2XF_1-8\dot{\phi}HF_2'=M_p^2-\frac{\xi\phi_0^2}{t_1^2}-16\frac{p\eta}{t_1^2},
\ee
is constant for the power-law solutions. For the case with $\phi_0=pM_p/4$ and  $p=20$ we find $M_p^{*2}\simeq M_p^2$. So we can conclude that, phenomenologically, the behavior described by the present model could still be a good asymptotic approximation since it comply with the current observations on dark energy and the propagation of GWs. 
It is worth noticing that in any case, the restriction $\alpha_g=0$ requires the restrictions on the parameters according to (\ref{eq31})-(\ref{eq33}) and additional tuning, including the fine tuning of the initial conditions for the scalar field is required if we want to maintain the relevance of the interaction terms at late times. \\
\noindent The limitations of this approach lie in the fact that $\dot{\phi}$, $\ddot{\phi}$ and $H$ are involved in the cancelation of the GW speed anomaly, making the approach valid only on the homogeneous FRW background, while this cancelation should take place in a covariant form, regardless of the underlying background. Otherwise, the perturbations of the background break the cancelation of the anomaly, leading to a residual contribution. \\
The above calculations were made for an scenario where the baryonic and dark matter are absent and the only matter content of the universe is in the form of dark energy.  
To approach a more realistic situation one should take into account the effect of matter. Here we will consider the matter term as an small contribution, and therefore, it will give a correction to the solutions (\ref{eq7}), (\ref{eq8}) by considering the effect of matter as a perturbation to the equations (\ref{eq3}) and (\ref{eq4}). Then we can evaluate the amount of correction that the account of matter gives to the restriction (\ref{eq25}). Expanding the equations (\ref{eq3}) and (\ref{eq4}) up to first order in the perturbations $\delta\phi$ and $\delta H$ around the solutions (\ref{eq7}), one finds 
 \be\label{eq3a}
 \begin{aligned}
\Big[\left(1+18p^2\xi_1\right)\phi_0 &+48p^3\frac{\eta_1}{\phi_0}\Big]\partial_t\delta\phi+\left[2\left(48p^3\eta_1-V_0t_1^2\right)\frac{1}{\phi_0}+18p^2\xi_1\phi_0\right]\frac{1}{t}\delta\phi +\\ &\Big[18p\xi_1\phi_0^2+144p^2\eta_1-6pM_p^2\Big]\delta H+\rho_{m} t=0
\end{aligned}
 \ee
 \be\label{eq4a}
 \begin{aligned}
&\left(1+6p^2\xi_1\right)\partial_t\partial_t\delta\phi+3p\left(1+6p^2\xi_1\right)\frac{1}{t}\partial_t\delta\phi+\\& 4\left[\left(V_0t_1^2+24p^3(p-1)\eta_1\right)\frac{1}{\phi_0^2}+3p^2(3p-2)\xi_1\right]\frac{1}{t^2}\delta\phi+12p\left[\xi_1\phi_0+4p\frac{\eta_1}{\phi_0}\right]\partial_t\delta H\\&
+3\phi_0\left[1+2p(9p-2)\xi_1+32p^2(2p-1)\frac{\eta_1}{\phi_0^2}\right]\frac{1}{t}\delta H=0
\end{aligned}
\ee 
where $\xi_1=\xi/t_1^2$ and $\eta_1=\eta/t_1^2$ and all functions have been evaluated on the power-law solution. We will assume  $\rho_m=\rho_{m0}a^{-3}$  in (\ref{eq3a}), which corresponds to dust matter that obeys the continuity equation. In order to integrate these equations and to have an estimate of the correction to the solutions (\ref{eq7}), we need the time behavior of the matter density $\rho_m$ . In the present case, given that the source of GWs is at very small redshift $z\sim 0.01$, one can make simplifications by taking the scale factor $a(t)$ to be constant during the time elapsed from the generation of the GW. Then we can write the matter contribution in the Eq. (\ref{eq3a}) as $\rho_m=3M_p^2H_0^2\Omega_{m0}$, where we used the normalization so that $a=1$. Nevertheless, after the integration of the Eqs. (\ref{eq3a}) and (\ref{eq3b}) it can be seen that $\delta\phi$ takes the form (using the expressions (\ref{eq31})-(\ref{eq33}) for $V_0$, $\xi_1$ and $\eta_1$)
\be\label{eq4b}
\delta\phi=f_1(\phi_0,p)t^2+c_1t^{\alpha_1(\phi_0,p)}+c_2t^{\alpha_2(\phi_0,p)}
\ee
where $c_1$ and $c_2$ are the integration constants and $f_1, \alpha_1, \alpha_2$ depend on $\phi_0/M_p$ and $p$ with very large number of terms.  Additionally, $\delta H$ contains a growing term $\propto t$. Despite the fact that one can make $f_1=0$ by fine tuning of $\phi_0$ and set $c_2=0$ (since $\alpha_2>0$), the term $\propto t$ in $\delta H$ is still present. This indicates that the power-law solution is unstable, under the assumption that the matter contribution remains constant during the time of propagation of the GW. Therefore the constant dark matter correction gives growing contribution to $\delta\alpha_g$, showing the impossibility of fixing the restriction $\alpha_g=0$ by fine tuning.  
Another approximation can be considered if one assumes that  $\rho_m\propto t^{-2}$, which is a solution for a fluid with constant equation of state in absence of other sources. Adding this matter term to the r.h.s of equation (\ref{eq3}), in this approximation the power-law solutions (\ref{eq7}) and (\ref{eq8}) still valid with the only restriction (\ref{eq9}) changed by (where $\rho_m=\left(\frac{3H_0^2t_1^2}{\kappa^2}\right)\Omega_{m0} t^{-2}$)
\be\label{eq9a}
\frac{3p^2}{\kappa^2}=\frac{1}{2}\phi_0^2+V_0 t_1^2+\frac{9\xi p^2}{t_1^2}\phi_0^2\pm \frac{48\eta p^3}{t_1^2}+\frac{3H_0^2t_1^2}{\kappa^2}\Omega_{m0},
\ee
and, therefore it will not affect the restriction (\ref{eq27}). The only change is that the matter density parameter enters in the expressions for $V_0, \xi, \eta$ as follows
\be\label{eq31a}
V_0=\frac{12(p^2-H_0^2t_1^2\Omega_{m0})(p\mp 1)^2M_p^2+(p^2\mp 6p+1)\phi_0^2}{2(2p^2\mp 3p-1)t_1^2}
\ee
\be\label{eq32a}
\xi=\frac{(p\mp 1)(2M_p^2(p^2-H_0^2t_1^2\Omega_{m0})\mp p\phi_0^2)t_1^2}{2p^2(2p^2\mp 3p-1)\phi_0^2}
\ee
\be\label{eq33a}
\eta=\frac{(\pm p\phi_0^2-2M_p^2(p^2-H_0^2t_1^2\Omega_{m0}))t_1^2}{16p^2(2p^2\mp 3p-1)}
\ee
the fact that the mater term evolves as it would do in the one-fluid universe, filled with matter with constant equation of state, is equivalent to the assumption that the matter density evolves proportional to the dark energy density (which is a situation known as scaling behavior valid for an earlier matter-dominated universe) which for the late time universe we consider as a naive approximation, since the rate of expansion is augmented by the increasingly dominant dark energy. But even if one accepts the solutions (\ref{eq7}) and (\ref{eq8}) as an approximation, one can analyze how the relative matter density fluctuations affect the power-law solutions. Under these fluctuations the scalar field and the Hubble parameter become (here we consider the solutions (\ref{eq7}))
$$\phi=\bar{\phi}+\delta\phi,\;\;\; H=\bar{H}+\delta H,$$ 
where $\bar{\phi}$ and $\bar{H}$ correspond now to the power-law solutions (\ref{eq7}) and any magnitude $f$ evaluated on the power-law solutions will be represented by $\bar{f}$.
For a perfect fluid with constant equation of state and density $\rho_m$, the continuity equation gives $\rho_m=\rho_{m0}a^{-3}$. Thus, the relative fluctuation of this density can be written as 
\be\label{eq3b}  
\frac{\delta\rho}{\rho}=-3\frac{\delta a}{a}=-3\delta N
\ee
where we introduced the slow-roll variable $N=\ln a$. On the other hand, one can connect the variation $\delta t$ with the density fluctuations as follows $$\delta t(N)=\frac{dt}{dN}\delta N=\frac{1}{H}\delta N,$$ which gives, from (\ref{eq7}) the variations
\be\label{eq3c}
\delta H=-\frac{1}{t}\delta N,\;\; \delta\phi=\frac{\phi_0}{p}\delta N,\;\; \partial_t\delta\phi=-\frac{\phi_0}{pt}\delta N,\;\; \partial_t\partial_t\delta\phi=\frac{2\phi_0}{pt^2}\delta N
\ee
where we used $\partial_t\delta N=\delta H.$
Then we can evaluate the amount of correction that $\delta\phi$ and $\delta H$ give to the restriction (\ref{eq25}), as
\be\label{eq3d}
\begin{aligned}
\delta{\alpha_g}\simeq &\frac{2}{M_p^{*2}}\Big[-4\bar{F}_2'\partial_t\partial_t\delta\phi+\left[2\dot{\bar{\phi}}\left(\bar{F}_1-4\bar{F}_2''\right)+4\bar{H}\bar{F}_2'\right]\partial_t\delta\phi+\\ &\left[\dot{\bar{\phi}}^2\left(\bar{F}_1'-4\bar{F}_2'''\right)-4(\ddot{\bar{\phi}}-\bar{H}\dot{\bar{\phi}})\bar{F}_2''\right]\delta\phi+4\dot{\bar{\phi}}\bar{F}_2'\delta H\Big]
\end{aligned}
\ee
where we kept the first order terms in $\delta\phi$ and $\delta H$. By using the basic power-law solutions (\ref{eq7}) (where the couplings $F_1$ and $F_2$ are given by (\ref{eq5})) in (\ref{eq3d}), it is obtained that $\delta\alpha_g=0$, even before using the restrictions (\ref{eq32}) and (\ref{eq33}) on $\xi$ and $\eta$ respectively. Then, by taking into account the quadratic terms in the expansion of $\alpha_g$, we find
\be\label{eq3e}
\delta\alpha_g\simeq-\frac{1}{\Big[1-\frac{\xi}{t_1^2}\frac{\phi_0^2}{M_p^2}-\frac{16\eta p}{M_p^2t_1^2}\Big]}\frac{16\eta}{M_p^2t_1^2 p}(\delta N)^2
\ee
Using the restrictions (\ref{eq32}) and (\ref{eq33}) we find
\be\label{eq3f}
\delta\alpha_g\simeq\frac{4p-2\gamma^2}{p\Big(4p^3-4p^2-(p+1)\gamma^2\Big)}(\delta N)^2
\ee
where $\phi_0$ is expressed in terms of the Planck mass as $\phi_0=\frac{2\sqrt{2}}{\alpha}M_p=\gamma M_p$. In this way we have found that, in the above qualitative analysis, $\alpha_g$ receives corrections of the order $(\delta N)^2$. Thus, if we assume that the relative matter density fluctuations, along the trajectory of the GW, are of the order of $\delta\rho/\rho\sim 10^{-1}$, then the correction to the velocity of the GW, given for instance $\gamma=p/4$ and $p=20$, becomes of the order of $\delta\alpha_g\sim 10^{-8}$ (taking into account that $\delta N=-\frac{1}{3}\delta\rho/\rho$), which is above the precision of the current measurements, and invalidates the restriction (\ref{eq26}). However from (\ref{eq3f}) it also follows that as the equation of state approaches the cosmological constant divide ($p\rightarrow\infty$), the correction $\delta\alpha_g\rightarrow 0$, which is consistent with the de Sitter solution. Note also that if one sets $\gamma=\sqrt{2p}$ (tuning $\phi_0$) in  (\ref{eq3f}), then $\delta\alpha_g=0$. In any case, (when it is possible) in order to keep $\delta\alpha_g$ either small or zero in presence of matter, additional tuning of the initial conditions is required.\\ 
The more practical way of solving the cosmological equations, taking into account the matter components, is by using the autonomous system technique. This method gives us the solutions as critical points where one can analyze the behavior of the fields, and at these points the scale factor acquires a power-law behavior. In the model with Gauss-Bonnet and non-minimal couplings \cite{granda10}, the are interesting stable critical points that are reached when the Gauss-Bonnet coupling is negligible and the dominant contribution comes from the non-minimal coupling. So, these results are not affected for the restrictions on the propagation of GW since the Gauss-Bonnet coupling becomes negligible. The same analysis is done in \cite{granda11}, where the model with kinetic and non-minimal couplings is considered. In this case there are also stable dark energy solutions at critical points where the kinetic coupling is negligible.  
This is consistent with the assumption that the restriction $\alpha_g=0$, takes place asymptotically in an scenario of dark energy dominance with the matter content "diluted" by the expansion rate of the universe.\\   
On the other hand, in the recent work \cite{weinberg} it has been shown that the effect of cold dark matter in the propagation of GWs, of nearby astrophysical origin (i.e. $z\lesssim 0.1$), is too small to be detected by current observations (see also \cite{gordon}). It was found that the effect of cold dark matter on damping and on modification of the propagation speed of GWs of astrophysical origin, was negligible. One should expect therefore, that the change in the dynamics of the scalar field due to the presence of cold dark matter could not affect perceptibly the restriction $\alpha_g=0$.\\
For the constraint $\alpha_g=0$ to be valid in any background, it should involve cancelation between scalars or tensor magnitudes of the same range. In the case of the present model the covariant cancellation of the GW speed anomaly (\ref{eq24}) imply 
$G_{4,X}=G_{5,\phi}$ and $G_{5,X}=0$, leading to the constraints $F_1=4F_2''$ and $F_2'=0$ respectively, but these constraints give rise to the standard uncoupled quintessence model from (\ref{eq1}). An alternative for the model (\ref{eq1}), in which the kinetic and GB couplings are retained, is to consider a combination of covariant and fine-tuning approach, which consists of keeping the constraint $F_1=4F_2''$ which is valid in any background and cancels the first two terms in the numerator of (\ref{eq24}), reducing the problem to fixing the restriction $G_{5,X}=0$ which affects the GB coupling. To deal with the restriction $XG_{5,X}\simeq 0$ we can appeal to the assumption that the cancellation of the defect in the speed of GWs is a local effect and try, for instance, the condition $F_2'(z= 0)=0$. to this end one could try the parametrization $F_2(z)\sim c_2 z^2+c_3z^3+...$\\
Alternatively we can apply arguments of fine-tuning, which we illustrate for the following case. If we assume the constraint
\be\label{eq34}
F_1=4F_2'',
\ee
which is valid in any background, we can use the following argument in order to perform the fine-tuning. 
Taking into account the above restriction (i.e. $F_1=4F_2''$), and assuming that over general backgrounds the expression for the GW speed anomaly $\alpha_g$, in the frame of the model (\ref{eq1}), can be written as
\be\label{eq36}
\alpha_g=\frac{8\mu^3 F_2'}{M_p^{*2}}
\ee
where $M_p^{*}$ is the effective Planck mass given by $M_p^{*2}=M_p^2-2XF_1+8\dot{\phi}HF_2'$, $\mu$ is a mass scale which could be interpreted as some upper bound on the coefficient of $G_{5,X}$ in (\ref{eq27}), and could be phenomenologically justified by cosmological bounds on $\dot{\phi}$ and $\ddot{\phi}$. Besides that, what we need is to establish an upper limit for $\alpha_g$. The fine-tuning is now tied to the behavior of the coupling function $F_2$. For the GB coupling of the form $\eta e^{-\gamma\kappa\phi}$ (not related to any specific solution), the decreasing exponential factor is crucial in the fine-tuning, allowing to turn the restriction $\alpha_g<10^{-16}$ into ($\kappa=M_p^{-1}$)
\be\label{eq37}
\frac{8\mu^3 \gamma\eta}{\sqrt{2}M_p M_p^{*2}}e^{-\frac{\gamma\phi}{M_p}}<10^{-16}.
\ee
A cosmologically relevant value for $\mu^3$ can be assumed as $\mu^3\sim M_p H_0^2$ and for the scalar field we can take the value $\phi\sim M_p$. Then, for late time universe the restriction becomes
\be\label{eq38}
\alpha_g\sim \frac{8H_0^2\gamma\eta}{M_p^{*2}}e^{-\gamma}<10^{-16}.
\ee
One can make a naive appreciation for $\alpha$ if we consider scenarios where $M_p^{*}$ is of the same order as $M_p$ and $\eta\sim M_p^2H_0^{-2}$, leading to $\gamma e^{-\gamma}<10^{-16}$, which gives  $\gamma>40$. The consequences for the GB interaction is that the expected contribution to the dark energy density, which should be of the order of $\sim M_p^2H_0^2$ becomes reduced by the factor $\gamma e^{-\gamma}\lesssim 10^{-16}$.
On the other hand, the contribution of the kinetic coupling becomes $F_1H_0^2\sim 4\frac{\eta\gamma^2H_0^2}{M_p^2}e^{-\gamma}\sim \gamma^2e^{-\gamma}$, which is a factor $\gamma$ larger than the GB contribution, but it's still very small contribution to dark energy.
Thus, the presence of the damping factor given by the decreasing exponent plays the role in satisfying the bound imposed by the GW170817 observations, but fails to give a relevant contribution to dark energy. Even if we consider an arbitrary coupling $F_2$, the fine-tuning falls on $F_2'$, producing the same effect on the contribution to dark energy.  So, the fine-tuning is always possible but at the cost of very tiny contribution to the dark energy.
\section{Discussion}
The recent discovery of gravitational waves from the NS merge in the system GW170817 imposes a severe bound on the velocity of propagation of the GW, and is currently of great importance to constraint the wide variety of models that have been considered as  serious and viable candidates for dark energy. As a consequence all models that make contribution to the anomalous velocity $c_g\ne c$ of GW (i.e. that give $\alpha_g\ne 0$), suffered severe restriction that drastically reduces the number of viable models of dark energy. This is the case of the generalized Galileons or Horndeski theory that accommodate most of the dark energy models proposed so far, including non-minimal derivative couplings to curvature and non-minimal coupling to the GB invariant. These last two interactions make the contribution to $\alpha_g$ given in (\ref{eq25}). However in the case of the scalar-tensor model (\ref{eq1}) with the exponential couplings (\ref{eq5}), the exponential can be factorized in the expression (\ref{eq25}) and the restriction $\alpha_g=0$, which is consistent with the observed value of $c_g$, leads to (\ref{eq26}). When the power-law solutions (\ref{eq7}) and (\ref{eq8}) are considered, we are left with the constraint (\ref{eq27}) which together with (\ref{eq9})-(\ref{eq11}) gives the solutions (\ref{eq31})-(\ref{eq33}).
The relevance of the couplings for the dark energy solution was shown for the case $\phi_0=pM_p/4$, $p=20$, where it was fount that the relative contribution to the density parameter, from the kinetic and GB terms is of the order of ${\cal O}(1)$. So, at least in the homogeneous FRW background and in the frame of the scalar-tensor model (\ref{eq1}), the cosmological solutions corresponding to power-law expansion, are a good approximation as they comply with the current observations on dark energy and the propagation of GWs. However it is worth noticing that in any case, the restriction $\alpha_g=0$ requires the restrictions on the parameters according to (\ref{eq31})-(\ref{eq33}) and additional tuning, including the fine tuning of the initial conditions is required if we want to maintain the relevance of the interaction terms at late times and comply with the dark energy observations.\\
It should be noted that the cancelation of the GW speed anomaly, in our case, takes place on an specific background while this cancelation should take place in a covariant from, regardless of the underlying background. Otherwise, the perturbations of the background break the cancelation, leading to a residual contribution which could be of second order effect, and here we assume that  this contribution is below the limit set by current observations. It is worth noting that the constraint $F_1=4F_2''$, valid in any background, cancels the terms  $G_{4,X}$ and $G_{5,\phi}$ in (\ref{eq24}), reducing the problem to the cancellation of $F_2'$. The only non-trivial ways to cancel $F_2'$ are the fine-tuning (that affects $F_1$ through the constraint), but as was shown in the example of exponential damping, the contribution of the couplings to the DE density could be reduced by a factor of $10^{-16}$. The other possibility is to consider that the cancellation of the GW speed anomaly is a local effect (the scale related to the emission and detection of GW170817 is much shorter than the cosmological scales relevant for DE), and in this case it might be  appropriate a parametrization of $F_2$ in terms of the redshift as proposed above. Resuming, we can conclude that the power-law cosmic accelerating expansion in the homogeneous FRW background can be driven by $G_4$ and $G_5$, generated by the non-minimal kinetic and GB couplings, while satisfying the restriction $\alpha_g=0$. The results are valid as long as the corrections to the defect in the velocity, caused by inhomogeneities, remain in the range below the current bounds on $\alpha_g$. 
\section*{Acknowledgments}
\noindent This work was supported by Universidad del Valle under project CI 71074 and by
COLCIENCIAS grant number 110671250405, DFJ acknowledges support from COLCIENCIAS,
Colombia.

\end{document}